\documentclass[aps,prl,twocolumn,groupedaddress,amsmath,amssymb,floatfix,superscriptaddress]{revtex4}
\usepackage{multirow}
\usepackage{graphicx}
\usepackage{color}
\usepackage{bm}
\begin{document}

\bibliographystyle{h-physrev3}

\title{Precision Analysis of the $^{136}$Xe Two-Neutrino $\beta\beta$ Spectrum in KamLAND-Zen \\ and Its Impact on the Quenching of Nuclear Matrix Elements}
%\title{Precision measurement of the $^{136}$Xe two-neutrino $\beta\beta$ spectrum in KamLAND-Zen \\ and its impact on the quenching of nuclear matrix elements}

% All university affiliations addresses go here:
\newcommand{\tohoku}{\affiliation{Research Center for Neutrino
    Science, Tohoku University, Sendai 980-8578, Japan}}
\newcommand{\ipmu}{\affiliation{Kavli Institute for the Physics and Mathematics of the Universe (WPI), 
    The University of Tokyo Institutes for Advanced Study, 
    The University of Tokyo, Kashiwa, Chiba 277-8583, Japan}}
\newcommand{\kyoto}{\affiliation{Kyoto University, Department of Physics, 
    Kyoto 606-8502, Japan}}
\newcommand{\osaka}{\affiliation{Graduate School of 
    Science, Osaka University, Toyonaka, Osaka 560-0043, Japan}}
\newcommand{\rcnp}{\affiliation{Research Center for Nuclear Physics, 
    Osaka University, Ibaraki, Osaka 567-0047, Japan}}
\newcommand{\tokushima}{\affiliation{Department of Physics, 
    Tokushima University, Tokushima 770-8506, Japan}}
\newcommand{\tokushimags}{\affiliation{Graduate School of Integrated Arts and Sciences, 
    Tokushima University, Tokushima 770-8502, Japan}}
\newcommand{\lbl}{\affiliation{Nuclear Science Division, Lawrence Berkeley National Laboratory,
    Berkeley, California 94720, USA}}
\newcommand{\hawaii}{\affiliation{Department of Physics and Astronomy,
    University of Hawaii at Manoa, Honolulu, Hawaii 96822, USA}}
\newcommand{\mitech}{\affiliation{Massachusetts Institute of Technology, 
    Cambridge, Massachusetts 02139, USA}}
\newcommand{\ut}{\affiliation{Department of Physics and
    Astronomy, University of Tennessee, Knoxville, Tennessee 37996, USA}}
\newcommand{\tunl}{\affiliation{Triangle Universities Nuclear Laboratory, Durham, 
    North Carolina 27708, USA; \\
    Physics Departments at Duke University, Durham, North Carolina 27708, USA; \\
    North Carolina Central University, Durham, North Carolina 27707, USA; \\
    and The University of North Carolina at Chapel Hill, Chapel Hill, North Carolina 27599, USA}}
\newcommand{\vt}{\affiliation{Center for Neutrino
   Physics, Virginia Polytechnic Institute and State University, Blacksburg,
   Virginia 24061, USA}}
\newcommand{\washington}{\affiliation{Center for Experimental Nuclear Physics and Astrophysics, 
    University of Washington, Seattle, Washington 98195, USA}}
\newcommand{\nikhef}{\affiliation{Nikhef and the University of Amsterdam, 
    Science Park, Amsterdam, the Netherlands}}
\newcommand{\tokyo}{\affiliation{Center for Nuclear Study, The University of Tokyo, 
    Tokyo 113-0033, Japan}}
\newcommand{\comenius}{\affiliation{Department of Nuclear Physics and Biophysics, 
    Comenius University, Mlynsk\'{a} dolina F1, SK-842 48 Bratislava, Slovakia}}
\newcommand{\dlnp}{\affiliation{Dzhelepov Laboratory of Nuclear Problems, 
    JINR 141980 Dubna, Russia}}
\newcommand{\bltp}{\affiliation{Bogoliubov Laboratory of Theoretical Physics, 
    JINR 141980 Dubna, Russia}}
\newcommand{\czech}{\affiliation{Czech Technical University in Prague, 
    128-00 Prague, Czech Republic}}

%
% Note: some authors have joint appointments with IPMU (http://www.ipmu.jp/members/)
%
% Tohoku
\author{A.~Gando}\tohoku
\author{Y.~Gando}\tohoku
\author{T.~Hachiya}\tohoku
\author{M.~Ha~Minh}\tohoku
\author{S.~Hayashida}\tohoku
\author{Y.~Honda}\tohoku
\author{K.~Hosokawa}\tohoku
\author{H.~Ikeda}\tohoku
\author{K.~Inoue}\tohoku\ipmu
\author{K.~Ishidoshiro}\tohoku
\author{Y.~Kamei}\tohoku
\author{K.~Kamizawa}\tohoku
\author{T.~Kinoshita}\tohoku
\author{M.~Koga}\tohoku\ipmu
\author{S.~Matsuda}\tohoku
\author{T.~Mitsui}\tohoku
\author{K.~Nakamura}\tohoku\ipmu
\author{A.~Ono}\tohoku
\author{N.~Ota}\tohoku
\author{S.~Otsuka}\tohoku
\author{H.~Ozaki}\tohoku
\author{Y.~Shibukawa}\tohoku
\author{I.~Shimizu}\tohoku
\author{Y.~Shirahata}\tohoku
\author{J.~Shirai}\tohoku
\author{T.~Sato}\tohoku
\author{K.~Soma}\tohoku
\author{A.~Suzuki}\tohoku
\author{A.~Takeuchi}\tohoku
\author{K.~Tamae}\tohoku
\author{K.~Ueshima}\tohoku
\author{H.~Watanabe}\tohoku

% IPMU
\author{D.~Chernyak}\ipmu
\author{A.~Kozlov}\ipmu

% Kyoto
\author{S.~Obara}\kyoto

% Osaka
\author{S.~Yoshida}\osaka

% RCNP
\author{Y.~Takemoto}\rcnp
\author{S.~Umehara}\rcnp

% Tokushima
\author{K.~Fushimi}\tokushima
\author{S.~Hirata}\tokushimags

% LBL
\author{B.E.~Berger}\ipmu\lbl
\author{B.K.~Fujikawa}\ipmu\lbl

% Hawaii
\author{J.G.~Learned}\hawaii
\author{J.~Maricic}\hawaii

% MIT
\author{L.A.~Winslow}\mitech

% UT
\author{Y.~Efremenko}\ipmu\ut

% TUNL
\author{H.J.~Karwowski}\tunl
\author{D.M.~Markoff}\tunl
\author{W.~Tornow}\ipmu\tunl

% VT
\author{T.~O'Donnell}\vt

% Washington
\author{J.A.~Detwiler}\ipmu\washington
\author{S.~Enomoto}\ipmu\washington

% NIKHEF
\author{M.P.~Decowski}\ipmu\nikhef

% Tokyo
\author{J.~Men\'endez}\tokyo

% Comenius
\author{R.~Dvornick\'y}\comenius\dlnp
\author{F.~\v{S}imkovic}\comenius\bltp\czech

\collaboration{KamLAND-Zen Collaboration}\noaffiliation

\date{\today}

\begin{abstract}
We present a precision analysis of the $^{136}$Xe two-neutrino $\beta\beta$ electron spectrum above 0.8\,MeV, based on high-statistics data obtained with the \mbox{KamLAND-Zen} experiment. An improved formalism for the two-neutrino $\beta\beta$ rate allows us to measure the ratio of the leading and subleading $2\nu\beta\beta$ nuclear matrix elements (NMEs), $\xi^{2\nu}_{31} = -0.26^{+0.31}_{-0.25}$. Theoretical predictions from the nuclear shell model and the majority of the quasiparticle random-phase approximation (QRPA) calculations are consistent with the experimental limit. However, part of the $\xi^{2\nu}_{31}$ range allowed by the QRPA is excluded by the present measurement at the 90\% confidence level. Our analysis reveals that predicted $\xi^{2\nu}_{31}$ values are sensitive to the quenching of NMEs and the competing contributions from low- and high-energy states in the intermediate nucleus. Because these aspects are also at play in neutrinoless $\beta\beta$ decay, $\xi^{2\nu}_{31}$ provides new insights toward reliable neutrinoless $\beta\beta$ NMEs.

\end{abstract}

\maketitle

{\it Introduction.}---Double-beta ($\beta\beta$) decay is a rare nuclear process. The $\beta\beta$ decay emitting two electron antineutrinos and two electrons ($2\nu\beta\beta$) is described within the standard model of the electroweak interaction. In contrast, the $\beta\beta$ mode without neutrino emission ($0\nu\beta\beta$) implies new physics, and can only occur if neutrinos are Majorana particles. While $2\nu\beta\beta$ decay has been measured in 12 isotopes~\cite{Barabash2015}, an observation of $0\nu\beta\beta$ decay remains elusive. In the standard scenario, the $0\nu\beta\beta$ rate is proportional to the square of the effective Majorana neutrino mass, $m_{\beta\beta}$~\cite{Avignone2008}, allowing the establishment of definite benchmarks toward the discovery of $0\nu\beta\beta$ decay in experiments.

The $0\nu\beta\beta$ rate, however, also depends on nuclear matrix elements (NMEs) which are poorly known~\cite{Engel2016}, as $0\nu\beta\beta$ NME estimates vary between the many-body approaches used to calculate them. In addition, NMEs may be affected by a possible ``quenching" or, equivalently, an effective value of the axial-vector coupling ${\it g}_{A}^{\rm eff}$ in the decay. Overall, the NME uncertainty can reduce the experimental sensitivity on $m_{\beta\beta}$ by up to a factor of 5~\cite{DellxOro2016}. To mitigate this, nuclear many-body predictions need to be tested in other observables. Several nuclear structure~\cite{Schiffer2008,Kay2009,Freeman2012,Brown2014} and Gamow-Teller (GT) properties~\cite{Yako2009,Frekers2013,Shimizu2018} have been proposed as $0\nu\beta\beta$ decay probes. Because $2\nu\beta\beta$ and $0\nu\beta\beta$ decays share initial and final nuclear states, and the transition operators are similar, a reproduction of $2\nu\beta\beta$ decay is key to reliable $0\nu\beta\beta$ NME predictions. Nonetheless, few nuclear many-body methods are well suited for both $\beta\beta$ modes, because nuclei with even and odd numbers of neutrons and protons up to high excitation energies need to be described consistently. The most notable approaches are the quasiparticle random-phase approximation (QRPA) ~\cite{Vogel1986,Engel1988,Suhonen1998,Simkovic2011,Simkovic2013} and the nuclear shell model~\cite{Caurier1990,Poves1995,Caurier2010,Horoi2013,Coraggio2017,Iwata2016}.

The $2\nu\beta\beta$ rate is usually expressed as
\begin{equation}
(T_{1/2}^{2\nu})^{-1} \simeq ({\it g}_{A}^{\rm eff})^{4} | M_{GT}^{2\nu} |^{2} G^{2\nu}_0,
\end{equation}
where $M_{GT}^{2\nu}$ is the $2\nu\beta\beta$ NME and $G^{2\nu}_0$ a known phase-space factor~\cite{Kotila2012}. As a result, ${\it g}_{A}^{\rm eff}$ can be determined from the measured $T_{1/2}^{2\nu}$ once $M_{GT}^{2\nu}$ is theoretically evaluated, a strategy followed in Ref.~\cite{ibm13}. While a similar approach has been used in the nuclear shell model, especially for $^{136}$Xe~\cite{Caurier2011,Horoi2013}, it is more common to take ${\it g}_{A}^{\rm eff}$ from GT $\beta$ decay and electron-capture (EC) rates~\cite{Caurier2011,Neacsu2015}, assuming a common quenching for all weak processes. Likewise, the QRPA can also use $\beta$ decay and EC to obtain ${\it g}_A^{\rm eff}$~\cite{lisi08,Suhonen2004,Suhonen2013}, even though the standard approach is to fix ${\it g}_{A}^{\rm eff}$ first, and then adjust the nuclear interaction so that $M_{GT}^{2\nu}$ describes the $2\nu\beta\beta$ half-life~\cite{Rodin2006}. In this way, the nuclear shell model and QRPA typically reproduce experimental $2\nu\beta\beta$ rates and predict nonmeasured ones~\cite{Caurier1990,Poves1995,Horoi2016,Pirinen2015,Perez2018}.

Recently, the $2\nu\beta\beta$ decay of several isotopes has been observed with high statistics by the \mbox{NEMO-3}~\cite{Arnold2015}, \mbox{EXO}~\cite{Albert2014a}, \mbox{KamLAND-Zen}~\cite{Gando2016}, \mbox{GERDA}~\cite{Agostini2017}, \mbox{\sc{Majorana}}~\cite{Aalseth2018} and \mbox{CUORE}~\cite{Alfonso2015} collaborations. These achievements demand an improved theoretical description. Reference~\cite{Simkovic2018} gives a more accurate expression for the $2\nu\beta\beta$ decay rate
\begin{align}
(T_{1/2}^{2\nu})^{-1} &\simeq  ({\it g}_{A}^{\rm eff})^{4} \left| (M_{GT}^{2\nu})^2 G_{0}^{2\nu}+M_{GT}^{2\nu}M_{GT-3}^{2\nu}\,G_{2}^{2\nu} \right| \nonumber \\
&=({\it g}_{A}^{\rm eff})^{4} | M_{GT-3}^{2\nu} |^{2} \frac{1}{|\xi_{31}^{2\nu}|^{2}} \left|G_{0}^{2\nu} + \xi_{31}^{2\nu} G_{2}^{2\nu}\right|,
\label{equation:2nu_rate}
\end{align} 
where the phase-space factor $G_{2}^{2\nu}$ has a different dependence on lepton energies than $G_{0}^{2\nu}$, and the subleading nuclear matrix element $M^{2\nu}_{GT-3}$ enters the (real-valued) ratio $\xi_{31}^{2\nu} = M_{GT-3}^{2\nu}/M_{GT}^{2\nu}$. While $M_{GT}^{2\nu}$ is sensitive to contributions from high-lying states in the intermediate odd-odd nucleus, for $M_{GT-3}$ only the lowest-energy states are relevant due to rapid suppression in the energy denominator. Consequently $\xi_{31}^{2\nu}$ probes additional, complementary physics to the $2\nu\beta\beta$ half-life. This novel observable can be determined experimentally by fitting the $2\nu\beta\beta$ electron energy spectrum to extract the leading and second order contributions in Eq.~(\ref{equation:2nu_rate}). Hence, the measurement of $\xi_{31}^{2\nu}$ challenges theoretical calculations and can discriminate between those that reproduce the $2\nu\beta\beta$ rate.

In this Letter, we analyze the high-statistics $2\nu\beta\beta$ decay of $^{136}$Xe with \mbox{KamLAND-Zen}~\cite{Gando2016} and compare the measured $T_{1/2}^{2\nu}$ and $\xi_{31}^{2\nu}$ values with the predictions from the QRPA and nuclear shell model. In \mbox{KamLAND-Zen}, the spectral distortion due to $\xi_{31}^{2\nu}$ could be up to 8\% based on the theoretical predictions. Such effect is testable with accumulated statistics of $\sim10^{5}$ $2\nu\beta\beta$ decays. Because $0\nu\beta\beta$ NMEs also show a competition between contributions from low- and high-energy intermediate states~\cite{Simkovic2011}, testing theoretical $\xi_{31}^{2\nu}$ predictions can provide new insights on $0\nu\beta\beta$ calculations, including the possible quenching of the NMEs.

{\it Experiment and results.}---The \mbox{KamLAND-Zen} (KamLAND Zero-Neutrino Double-Beta Decay)  detector consists of \mbox{13\,tons} of Xe-loaded liquid scintillator~(\mbox{Xe-LS}) contained in a 3.08-m-diameter spherical inner balloon (IB). The IB is constructed from 25-$\mu$m-thick transparent nylon film and is suspended at the center of the KamLAND detector~\cite{Gando2015,Gando2013b}. The IB is surrounded by 1\,kton of liquid scintillator (LS) which acts as an active shield. The scintillation photons are viewed by 1879 photomultiplier tubes mounted on the inner surface of the containment vessel. The \mbox{Xe-LS} consists of 80.7\% decane and 19.3\% pseudocumene (1,2,4-trimethylbenzene) by volume, 2.29\,g/L of the fluor PPO (2,5-diphenyloxazole), and $(2.91 \pm 0.04)$\% by weight of enriched xenon gas. The isotopic abundances in the enriched xenon were measured by a residual gas analyzer to be $(90.77 \pm 0.08)\%$ \mbox{$^{136}$Xe}, $(8.96 \pm 0.02)\%$ \mbox{$^{134}$Xe}. 

We report on data collected between December 11, 2013 and October 27, 2015, which is the same data set analyzed for the $0\nu\beta\beta$ search in Ref.~\cite{Gando2016} with a total live time of 534.5~days. The selection to reduce the background contributions is the same as in Ref.~\cite{Gando2016}, but we apply a tightened $2\nu\beta\beta$ event selection for this work in order to avoid systematic uncertainties arising from backgrounds. The fiducial volume for the reconstructed event vertices is defined as a 1-m-radius spherical shape at the detector center, which gives a fiducial exposure for this analysis of $(126.3 \pm 3.9)$\,kg yr in $^{136}$Xe. We perform a likelihood fit to the binned energy spectrum of the selected candidates between 0.8 and 4.8\,MeV, tightened relative to the $2\nu\beta\beta$ analysis in Ref.~\cite{Gando2016}. The systematic uncertainties on the $2\nu\beta\beta$ rate are evaluated identically as in Ref.~\cite{Gando2016} and are summarized in Table~\ref{table:systematic}.

\begin{center}
\begin{table}[b]
\caption{\label{table:systematic}Estimated systematic uncertainties used for the $^{136}$Xe $2\nu\beta\beta$ decay rate measurement.
}
\begin{tabular}{@{}*{2}{lc}}
\hline
\hline
Source  \hspace{3.5cm} & Systematic uncertainty (\%) \\
\hline
Fiducial volume & 3.0 \\
Enrichment factor of $^{136}$Xe & 0.09 \\
Xenon mass & 0.8 \\
Detector energy scale & 0.3 \\
Detection efficiency & 0.2 \\
\hline
Total & 3.1 \\
\hline
\hline
\end{tabular}
\end{table}
\end{center}

\begin{figure}[t]
	\includegraphics[width=1.0\columnwidth]{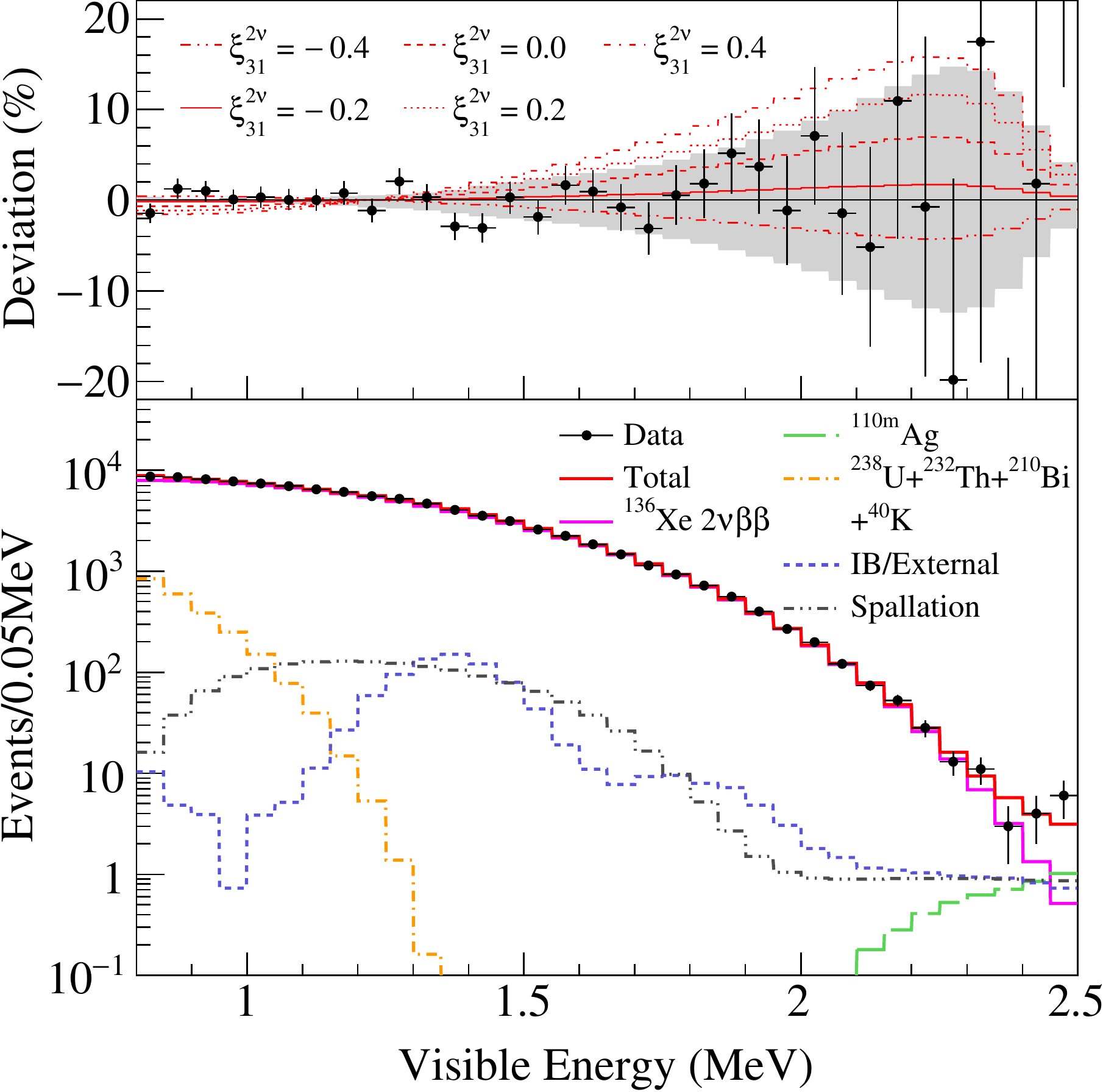}
	\caption{Bottom panel: Observed energy spectrum of selected $2\nu\beta\beta$ candidates within a 1-m-radius spherical volume (dotted) drawn together with best-fit backgrounds and the $2\nu\beta\beta$ decay spectrum floating the value of $\xi_{31}^{2\nu}$. Top panel: Deviation of the observed spectrum (dotted) from the best-fit ($\xi_{31}^{2\nu} = -0.26$). The lines indicate the expectation for $\xi_{31}^{2\nu} = -0.4, -0.2, 0.0, 0.2, 0.4$. The shaded band represents the systematic uncertainty due to the energy scale error.}
	\vspace{-0.3cm}
	\label{figure:energy_spectrum}
\end{figure}

A detailed energy calibration is essential for the extraction of $\xi_{31}^{2\nu}$. The energy scale was determined using $\gamma$ rays from $^{60}$Co, $^{68}$Ge, and $^{137}$Cs radioactive sources, $\gamma$ rays from the capture of spallation neutrons on protons and $^{12}$C, and $\beta + \gamma$-ray emissions from $^{214}$Bi, a daughter of $^{222}$Rn (lifetime 5.5\,day) that was introduced during the \mbox{Xe-LS} purification. Uncertainties from the nonlinear energy response due to scintillator quenching and Cherenkov light production are constrained by the calibrations. The most important calibration is the high-statistics $^{214}$Bi from the initial $^{222}$Rn distributed uniformly over the \mbox{Xe-LS} volume. To ensure that the calibration with $^{214}$Bi can be applied to the entire data set, we confirmed that the time variation of the energy scale is less than 0.5\% based on the spectral fit to the $2\nu\beta\beta$ decays for each time period. This uncertainty is reduced relative to the previous analysis~\cite{Gando2016}, and is added to the energy scale error, which is the dominant error source for the $\xi_{31}^{2\nu}$ measurement, as discussed later.

The energy spectrum of selected candidate events between 0.8 and 2.5\,MeV together with the best-fit spectral decomposition is shown in Fig.~\ref{figure:energy_spectrum}. In the fit, the contributions from $2\nu\beta\beta$ and major backgrounds in the \mbox{Xe-LS}, such as $^{40}$K, $^{210}$Bi, and the $^{228}$Th-$^{208}$Pb subchain of the $^{232}$Th series are free parameters and are left unconstrained. The background contribution from $^{110m}$Ag, which is important for the $0\nu\beta\beta$ analysis, is also a free parameter in the fit. The contributions from the $^{222}$Rn-$^{210}$Pb subchain of the $^{238}$U series, and from $^{11}$C and $^{10}$C (muon spallation products), as well as the detector energy response model parameters, are allowed to vary but are constrained by their independent estimations~\cite{Gando2016}.

\begin{figure}[t]
\includegraphics[width=1.0\columnwidth]{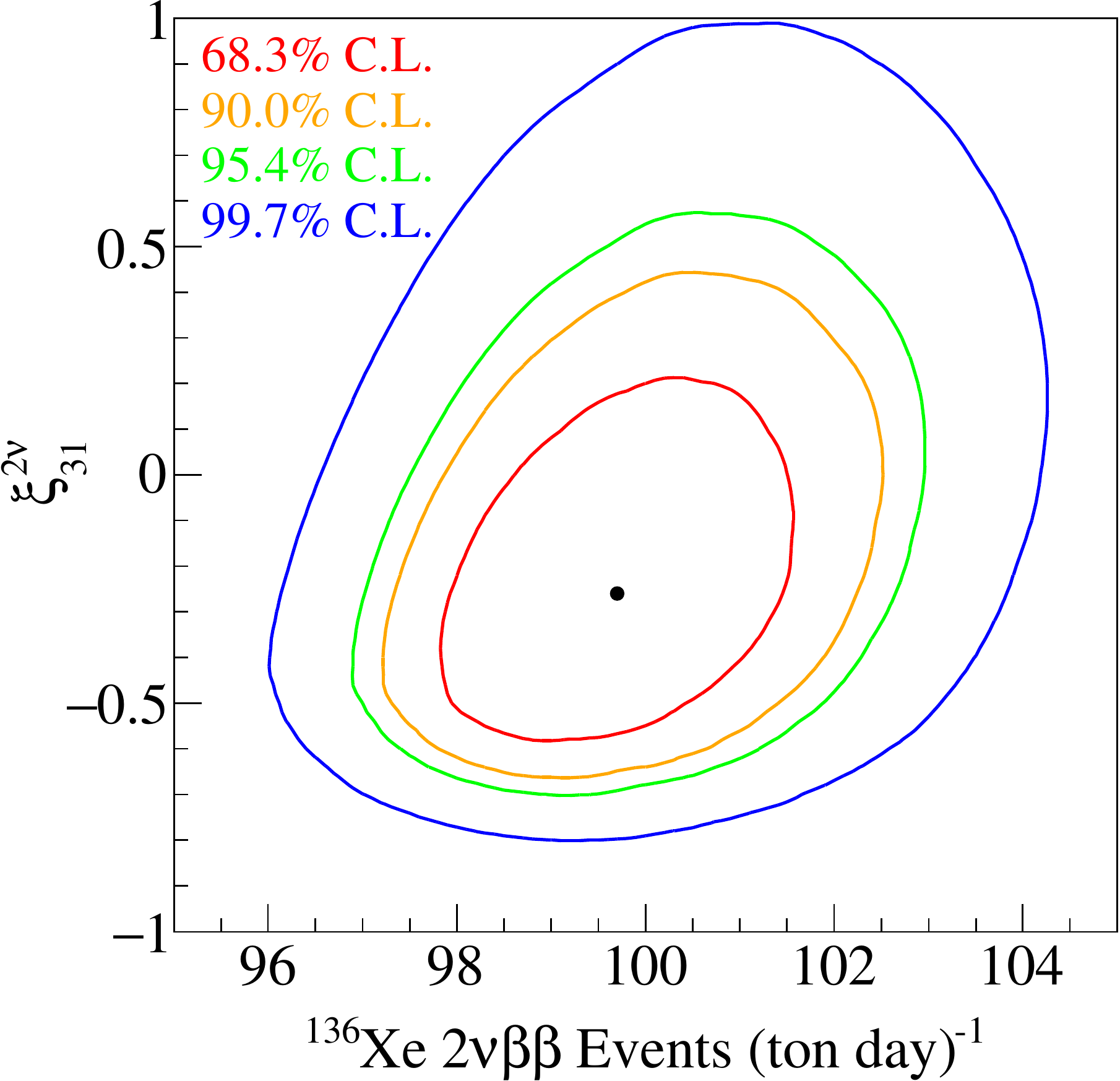}
  \caption{Allowed region for the joint variation of the $^{136}$Xe $2\nu\beta\beta$ decay rate and the ratio of the matrix elements $\xi_{31}^{2\nu}$ at the 68.3\%, 90\%, 95.4\%, and 99.7\% confidence levels (C.L.). The dot represents the best-fit point. The profile for $\xi_{31}^{2\nu}$ gives a best-fit of $\xi_{31}^{2\nu} = -0.26^{+0.31}_{-0.25}$ and a 90\% C.L. upper limit of $\xi_{31}^{2\nu} < 0.26$.}
  \vspace{-0.3cm}
  \label{figure:contour}
\end{figure}

The $2\nu\beta\beta$ spectrum is computed with Eq.~(\ref{equation:2nu_rate}), convolved with the detector response function. It is characterized by two free parameters: the total $2\nu\beta\beta$ rate and the ratio of the matrix elements $\xi_{31}^{2\nu}$. We obtained a best fit of $\xi_{31}^{2\nu} = -0.26^{+0.31}_{-0.25}$ and a 90\% C.L. upper limit of $\xi_{31}^{2\nu} < 0.26$. The systematic uncertainty on the energy scale limits the sensitivity of the $\xi_{31}^{2\nu}$ measurement, because an energy scale shift introduces a shape distortion similar to the change generated by a nonzero $\xi_{31}^{2\nu}$. The best-fit total $2\nu\beta\beta$ rate in the Xe-LS mass is $99.7^{+1.2}_{-1.4}$\,\mbox{(ton day)$^{-1}$}. Figure~\ref{figure:contour} shows the joint confidence intervals for the $2\nu\beta\beta$ rate and $\xi_{31}^{2\nu}$, which exhibit only a slight positive correlation. It indicates that the effect on the total $2\nu\beta\beta$ rate estimate by the introduction of the second order contribution is small. The effect on the $0\nu\beta\beta$ analysis is also negligibly small. Considering the systematic uncertainties in Table~\ref{table:systematic}, the $2\nu\beta\beta$ decay half-life of $^{136}$Xe is estimated to be $T_{1/2}^{2\nu} = 2.23 \pm 0.03({\rm stat}) \pm 0.07({\rm syst}) \times 10^{21}$\,yr. This result is consistent with our previous result based on \mbox{phase-II} data, $T_{1/2}^{2\nu} = 2.21 \pm 0.02({\rm stat}) \pm 0.07({\rm syst}) \times 10^{21}$\,yr~\cite{Gando2016}, and with the result obtained by \mbox{EXO-200}, $T_{1/2}^{2\nu} = 2.165 \pm 0.016({\rm stat}) \pm 0.059({\rm syst}) \times 10^{21}$\,yr~\cite{Albert2014a}. Our analysis neglects counts from decays to excited states in $^{136}$Ba, for which our shell model calculations predict $T_{1/2}^{2\nu} (0_{\rm gs}^{+} \rightarrow 0_{1}^{+}) > 10^{26}$\,yr. Even a very conservative half-life of $8.7 \times 10^{24}$\,yr that assumes the same NME for the decay to the ground (gs) and excited $0^{+}$ states does not affect our results. This issue might need to be revisited in the case of an unexpectedly short half-life close to the present 90\% C.L. lower limit of $8.3 \times 10^{23}$\,yr~\cite{Asakura2016}. The correction to $2\nu\beta\beta$ decay represented by $\xi_{31}^{2\nu}$ impacts \mbox{KamLAND-Zen} analyses of spectral distortions, including extraction of half-lives to excited states as well as searches for beyond-standard-model physics, such as for Majoron emission modes. Considering $\xi_{31}^{2\nu}$ as a free parameter, we find the additional uncertainty comparable to the energy scale error. Updated spectral analyses will be presented in future publications.

{\it Theoretical calculations.}---We obtain the $2\nu\beta\beta$ decay NMEs $M_{GT}^{2\nu}$ and $M_{GT-3}^{2\nu}$ to compare calculated $\xi^{2\nu}_{31}$ values to the \mbox{KamLAND-Zen} limit. The NMEs are defined as~\cite{Simkovic2018}
\begin{align}
M_{GT}^{2\nu}&=
\sum_j\frac{\langle 0^+_{f} | \sum_l {\bm \sigma}_l \tau^-_l | 1^+_j \rangle
	\langle 1^+_j | \sum_l {\bm \sigma}_l \tau^-_l | 0^+_{i}\rangle}
{\Delta} \,,
\label{M2n} \\
M_{GT-3}^{2\nu}&=\!\sum_j\frac{4\langle 0^+_{f} | \sum_l {\bm \sigma}_l \tau^-_l | 1^+_j \rangle
	\langle 1^+_j | \sum_l {\bm \sigma}_l \tau^-_l | 0^+_{i}\rangle}
{\Delta^3},
\label{M2n3}
\end{align}
with energy denominator $\Delta=[E_j-(E_i+E_f)/2]/m_e$. $E_k$ is the energy of the nuclear state $|J^\pi_{k}\rangle$ with total angular momentum $J$ and parity $\pi$, and $m_e$ is the electron mass. The labels $i$, $j$, $f$ refer to the initial, intermediate and final nuclear states, respectively, while ${\bm \sigma}$ is the spin and $\tau^-$ the isospin lowering operator.

We perform nuclear shell model calculations in the configuration space comprising the $0g_{7/2}$, $1d_{5/2}$, $1d_{3/2}$, $2s_{1/2}$, and $0h_{11/2}$ single-particle orbitals for both neutrons and protons, using the shell model code NATHAN~\cite{Caurier2004}. We reproduce $M_{GT}^{2\nu}=0.064$ from Ref.~\cite{Caurier2011} with the GCN interaction~\cite{Caurier2010}, and also use the alternative MC interaction from Ref.~\cite{Qi2012}, which yields $M_{GT}^{2\nu}=0.024$. Both interactions have been used in $0\nu\beta\beta$ decay studies~\cite{Menendez2009,Shimizu2018}. Shell model NMEs for $\beta$ and $2\nu\beta\beta$ decays are typically too large, due to a combination of missing correlations beyond the configuration space, and neglected two-body currents in the transition operator~\cite{Engel2016}. This is phenomenologically corrected with a ``quenching" factor $q$, or ${\it g}_A^{\rm eff}=q\,{\it g}_A$. In general, the quenching that fits GT $\beta$ decays and ECs in the same mass region is valid for $2\nu\beta\beta$ decays as well. Around $^{136}$Xe, GT transitions with GCN are best fit with $q=0.57$~\cite{Caurier2011}, and with the same adjustment the $^{136}$Xe GT strength into $^{136}$Cs~\cite{Frekers2013}, available up to energy $E\lesssim4.5$\,MeV, is well reproduced by both interactions. However, the experimental $2\nu\beta\beta$ half-life suggests different quenching factors $q=0.42(0.68)$ for GCN (MC). The calculations yield $M_{GT-3}^{2\nu}=0.011(0.0025)$. We assume a common quenching for $M_{GT}^{2\nu}$ and $M_{GT-3}^{2\nu}$ because the shell model reproduces well GT strengths at low and high energies up to the GT resonance~\cite{Yako2009}. This gives ratios $\xi^{2\nu}_{31}=0.17$ for GCN and $\xi^{2\nu}_{31}=0.10$ for MC, both consistent with the present experimental analysis.

\begin{figure}[t]
	\begin{center}
		\includegraphics[width=1.0\columnwidth]{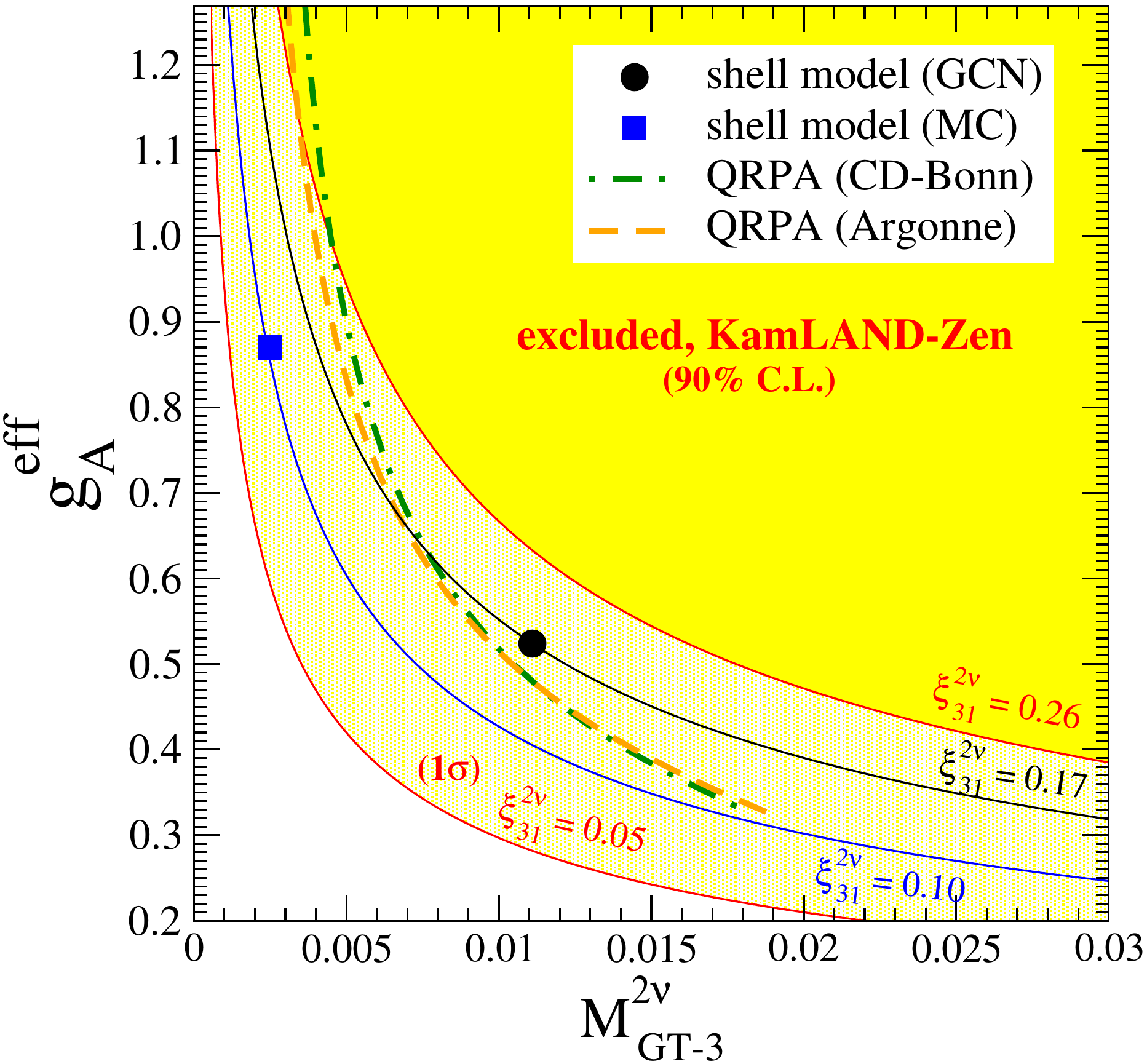}
	\end{center}
	\caption{Effective axial-vector coupling ${\it g}_A^{\rm eff}$
		as a function of the matrix element $M^{2\nu}_{GT-3}$ for
		$^{136}$Xe $2\nu\beta\beta$ decay. The yellow (light yellow) region $\xi^{2\nu}_{31} > 0.26$ (0.05) is excluded by the present \mbox{KamLAND-Zen} measurement at 90\% (1$\sigma$) C.L.
		Nuclear shell model results are displayed by
		the black circle (GCN interaction) and blue square (MC).
		QRPA results are shown by the dashed orange (Argonne interaction) and dashed-dotted green (CD-Bonn) curves.
		\label{fig.gA}}
\end{figure}

We also perform $2\nu\beta\beta$ decay QRPA calculations with partial restoration of isospin symmetry \cite{Simkovic2013}. We consider a configuration space of 23 single-particle orbitals (the six lowest harmonic oscillator shells with the addition of the $0i_{13/2}$ and $0i_{11/2}$ orbitals). We take as nuclear interactions two different G matrices, based on the charge-dependent Bonn (CD-Bonn) and the Argonne V18 nucleon-nucleon potentials. We fix the isovector proton-neutron interaction imposing the restoration of isospin~\cite{Simkovic2013}. Finally, we adjust the isoscalar neutron-proton interaction to reproduce the $2\nu\beta\beta$ decay half-life for different values in the range ${\it g}_A^{\rm eff} \le {\it g}_A=1.269$. We obtain the following ranges of results: $M_{GT}^{2\nu}=(0.011,0.164)$, $M_{GT-3}^{2\nu}=(0.0031,0.019)$ and $\xi^{2\nu}_{31}=(0.11, 0.29)$ for the Argonne potential; and $M_{GT}^{2\nu}=(0.011,0.157)$, $M_{GT-3}^{2\nu}=(0.0036,0.018)$ and $\xi^{2\nu}_{31}=(0.11, 0.35)$ using the CD-Bonn potential. Except for the larger $\xi^{2\nu}_{31}$ values, especially with CD-Bonn,
most of the QRPA predictions are consistent with the present experimental analysis.

{\it Discussion.}---Figure~\ref{fig.gA} shows the effective axial-vector coupling constant ${\it g}_A^{\rm eff}$ as a function of the matrix element $M^{2\nu}_{GT-3}$ for the $2\nu\beta\beta$ decay of $^{136}$Xe. A large region in the ${\it g}_A^{\rm eff}-M^{2\nu}_{GT-3}$ plane is excluded by the present 90\% C.L. limit $\xi^{2\nu}_{31} < 0.26$. The two nuclear shell model GCN and MC results, indicated by points, are consistent with the \mbox{KamLAND-Zen} limit. The QRPA Argonne and CD-Bonn results are presented by curves, which accommodate $0.33 \leq {\it g}_A^{\rm eff} \leq 1.269$ values (the lower end corresponds to vanishing isoscalar interactions). Both curves are very similar, because QRPA ratios of matrix elements with the same initial and final states are weakly sensitive to the nucleon-nucleon interaction~\cite{Rodin2006}. Figure~\ref{fig.gA} shows that, even though most QRPA predictions are consistent with our measurement, ${\it g}_A^{\rm eff}\gtrsim1.14(1.00)$ for the Argonne (CD-Bonn) potential is excluded at 90\% C.L. by the \mbox{KamLAND-Zen} $\xi^{2\nu}_{31}$ limit.

Figure~\ref{fig.gA} also shows that for ${\it g}_A^{\rm eff}\gtrsim 0.7$ the QRPA predicts larger $\xi^{2\nu}_{31}$ values than the nuclear shell model. Elsewhere, the QRPA ratios lie between those of the GCN and MC shell model interactions. Interestingly, for ${\it g}_A^{\rm eff}\sim0.5$, the QRPA and shell model GCN results are close. While such relatively small ${\it g}_A^{\rm eff}$ values are not always considered in $2\nu\beta\beta$ QRPA calculations of $^{136}$Xe, they are favored by QRPA statistical analyses that take into account experimental EC and $\beta$ rates~\cite{lisi08,deppisch16}.

\begin{figure}[t]
	\includegraphics[width=1.0\columnwidth]{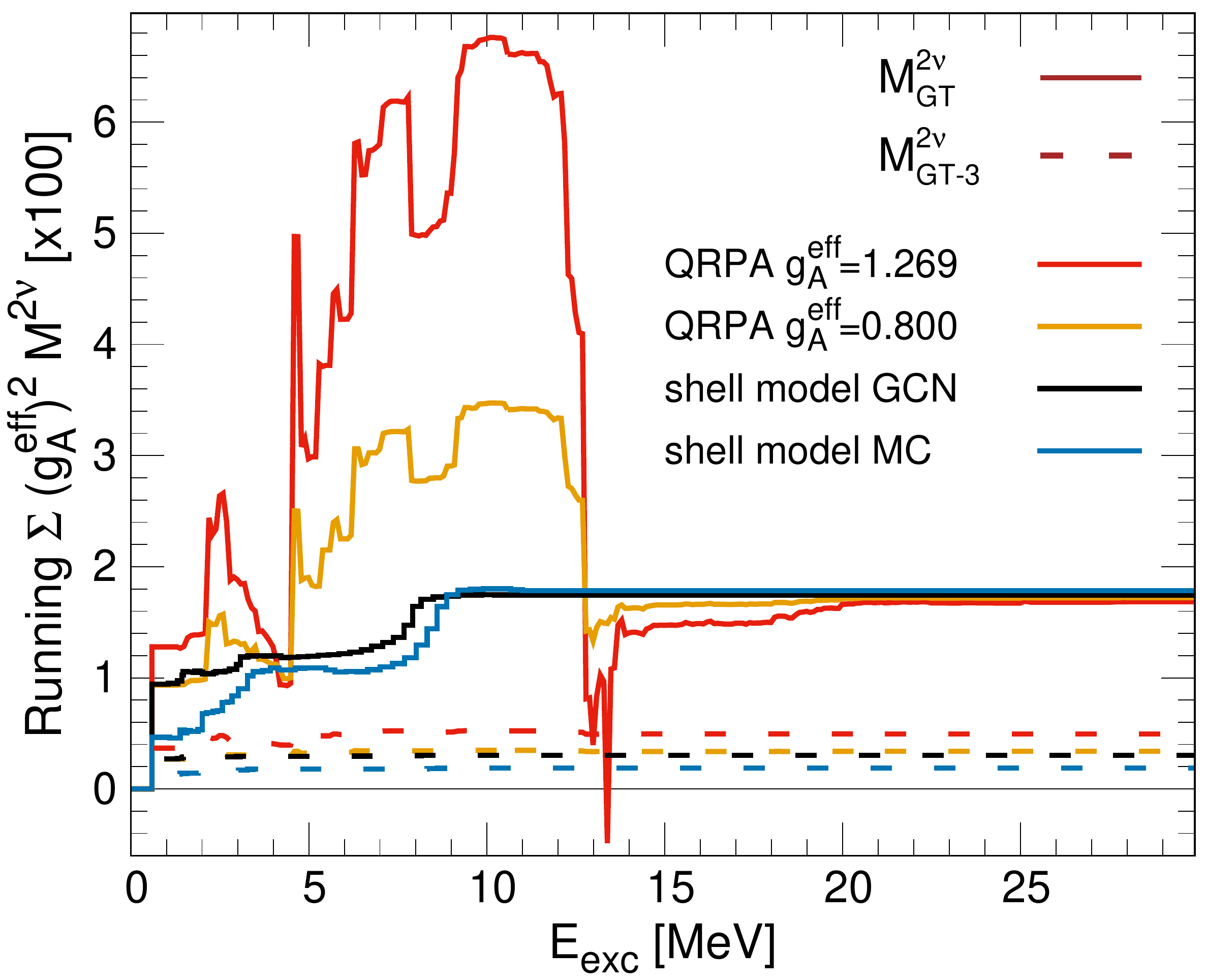}
	\caption{Running sum of the $^{136}$Xe $M_{GT}^{2\nu}$ (solid lines) and $M_{GT-3}^{2\nu}$ (dashed) $2\nu\beta\beta$ NMEs, as a function of the excitation energy of the $1^+$ states in $^{136}$Cs. Nuclear shell model results with the GCN (MC) interaction, indicated by black (blue) lines, are compared to the QRPA Argonne running sum with ${\it g}_A^{\rm eff}=1.269$ (${\it g}_A^{\rm eff}=0.80$), shown by red (orange) lines.
	}
	\vspace{-0.3cm}
	\label{figure:M2n_running}
\end{figure}

To illustrate the origin of the differences between the theoretical calculations, Fig.~\ref{figure:M2n_running} compares the nuclear shell model and QRPA Argonne running sums of $M_{GT}^{2\nu}$ and $M_{GT-3}^{2\nu}$~\cite{Simkovic2018,Caurier2011}, multiplied by the corresponding $({\it g}_A^{\rm eff})^{2}$. The sums run over the excitation energy of the spin-parity $1^+$ states in the intermediate nucleus $^{136}$Cs. The theoretical $M_{GT}^{2\nu}$ running sums differ: while the shell model converges at $E_{\rm exc}\simeq8$~MeV, QRPA terms contribute until $E_{\rm exc}\simeq20$~MeV. Moreover, at $E_{\rm exc}\sim10$~MeV the accumulated QRPA $M_{GT}^{2\nu}$ exceeds the shell model significantly, with a strong ${\it g}_A^{\rm eff}$ sensitivity. While for ${\it g}_A^{\rm eff} = 1.269$ the maximum of the QRPA running sum is almost four times larger than the shell model one, for ${\it g}_A^{\rm eff} \sim 0.5$ ---not shown in Fig.~\ref{figure:M2n_running}--- the difference is only about 20\%, consistent with the more similar $\xi^{2\nu}_{31}$ values predicted. $E_{\rm exc}\sim10$~MeV shell-model contributions may be too small due to missing spin-orbit partner orbitals, but the QRPA may also overestimate them. Measurements of charge-exchange reactions up to the $^{136}$Xe GT resonance, currently limited to lower energy~\cite{Frekers2013,Puppe2011}, can clarify this picture. Above $E_{\rm exc}\gtrsim10$~MeV, the QRPA excess with respect to the shell model is canceled. The final value, set by the $2\nu\beta\beta$ half-life, is common to all calculations.

By contrast, Fig.~\ref{figure:M2n_running} shows that in both shell model and QRPA the lowest $1^+$ state component dominates the $M_{GT-3}^{2\nu}$ NME. Such contribution is more salient for the shell model GCN and QRPA ${\it g}_A^{\rm eff}=1.269$ calculations, which explains the larger associated $\xi^{2\nu}_{31}$ value compared to the shell model MC and QRPA ${\it g}_A^{\rm eff}=0.8$ results, respectively. The contrast in the $M_{GT-3}^{2\nu}$ running sum at low energies is ultimately responsible for the different $\xi^{2\nu}_{31}$ values predicted by the QRPA and nuclear shell model.

In $0\nu\beta\beta$ decay, the running sum of the NME can extend to even higher energies, because in this case there is no dependence on the energy of the intermediate states in the denominator; see Eqs.~(\ref{M2n}) and (\ref{M2n3}). Therefore, a competition between contributions from low- and high-energy states similar to $2\nu\beta\beta$ decay is expected~\cite{Simkovic2011,Senkov2013,Senkov2014}. Consequently, fixing $\xi^{2\nu}_{31}$ in $2\nu\beta\beta$ decay will allow one to identify the most promising $0\nu\beta\beta$ NME predictions.

Further experimental $\xi^{2\nu}_{31}$ sensitivity improvements may distinguish between various scenarios. On the one hand, measured values of $\xi^{2\nu}_{31} \ge 0.11$ will allow QRPA calculations to fix the quenched value of ${\it g}_A^{\rm eff}$, reducing uncertainties in QRPA $0\nu\beta\beta$ NMEs~\cite{Rodin2006,Ejiri2019,Terasaki2016}. Likewise, a measured value $\xi^{2\nu}_{31}\simeq0.17 (0.10)$ would suggest that the GCN (MC) shell model interaction, with its associated ${\it g}_A^{\rm eff}$ value, leads to a more reliable $0\nu\beta\beta$ NME. Because the QRPA and shell model rely on different assumptions, and for $2\nu\beta\beta$ decay they can exhibit contrasting sensitivities on ${\it g}_A^{\rm eff}$ ---as shown in Fig.~\ref{figure:M2n_running}--- a measurement of $\xi^{2\nu}_{31}$ could lead to different ${\it g}_A^{\rm eff}$ values for each model. Furthermore, the quenching may not be the same in $2\nu\beta\beta$ and $0\nu\beta\beta$ decays, especially in the light of the differences in the two-body~\cite{Menendez2011,Engel2014,Wang2018} and contact~\cite{Cirigliano2018} corrections to the two $\beta\beta$ transition operators. On the other hand, a small ratio $\xi^{2\nu}_{31} < 0.11$, which cannot be accommodated in the present QRPA calculations, or a determination of $\xi^{2\nu}_{31}$ very different to the GCN and MC predictions, would demand improved theoretical developments.

{\it Summary.}---We have presented a precision analysis of the $^{136}$Xe $2\nu\beta\beta$ electron spectrum shape with the \mbox{KamLAND-Zen} experiment. For the first time, we set a limit on the ratio of nuclear matrix elements $\xi^{2\nu}_{31}<0.26$ (90\% C.L.). The experimental limit is consistent with the predictions from the nuclear shell model and most QRPA calculations, but excludes QRPA Argonne (CD-Bonn) results for ${\it g}_A^{\rm eff}\gtrsim$ 1.14(1.00). The allowed theoretical values vary in the range $\xi^{2\nu}_{31}=(0.10-0.26)$, so that future $\xi^{2\nu}_{31}$ measurements will be required to further test $2\nu\beta\beta$ calculations, and select the most successful ones. The associated ${\it g}_A^{\rm eff}$ value, or NME quenching, would also be identified. Future experiments such as \mbox{KamLAND2-Zen}~\cite{Inoue2013} and others with improved resolution and reduced backgrounds promise enhanced sensitivity to reach this goal. Our analysis reveals that $\xi^{2\nu}_{31}$ is sensitive to competing contributions to the NME from low- and high-energy intermediate states. Because a similar competition is also relevant for $0\nu\beta\beta$ decay, studies of this observable provide new insights for identifying reliable $0\nu\beta\beta$ NMEs.

\vspace{1cm}
\begin{acknowledgments}
We thank P. Vogel for useful discussions. The \mbox{KamLAND-Zen} experiment is supported by JSPS KAKENHI Grants No. 21000001 and 26104002; the World Premier International Research Center Initiative (WPI Initiative), MEXT, Japan; Netherlands Organisation for Scientific Research; and under the U.S. Department of Energy (DOE) Award No.\,DE-AC02-05CH11231, as well as other DOE and NSF grants to individual institutions. The Kamioka Mining and Smelting Company has provided service for activities in the mine. We acknowledge the support of NII for SINET4. J.M. is supported by the JSPS KAKENHI Grant No. 18K03639, MEXT as "Priority issue on post-K computer" (Elucidation of the fundamental laws and evolution of the universe), JICFuS, the CNS-RIKEN joint project for large-scale nuclear structure calculations, and the U.S. DOE (Award No. DE-FG02-00ER41132). J.M. thanks the Institute for Nuclear Theory at the University of Washington for its hospitality. F. S. is supported by the Slovak Research and Development Agency under Contract No. APVV-14-0524 and European Regional Development Fund-Project "Engineering applications of microworld physics" 
(CZ.02.1.01/0.0/0.0/16\_019/0000766).
\end{acknowledgments}

\bibliography{DoubleBetaAxialVector}

\end{document}